\documentclass[epj]{svjour}
\usepackage[T1]{fontenc}

\usepackage[utf8x]{inputenc}
\usepackage{graphicx}
\usepackage{mathptmx}      
\usepackage{flushend}
\usepackage{fancyvrb}
\usepackage[colorlinks,citecolor=blue,urlcolor=blue,linkcolor=blue]{hyperref}

\newcommand{\ttH}{$t\bar{t}H$\space}
\newcommand{\ttbar}{t\bar{t}\space}

\begin{document}

\title{HEP-Frame: an Efficient Tool for Big Data Applications at the LHC\footnote{The research leading to these results was partially funded by Fundacao para a Ciencia e Tecnologia under Grant Agreement No UIDB/00319/2020. The authors have no competing interests to declare that are relevant to the content of this article.}}

\author{André Pereira\inst{1}\thanks{ORCID: 0000-0002-2110-914X, ampereira@macc.fccn.pt - Corresponding Author}
        \and
        António Onofre\inst{2}\thanks{ORCID: 0000-0003-3471-2703, antonio.onofre@cern.ch}
        \and
        Alberto Proença\inst{3}\thanks{ORCID: 0000-0001-6018-7346, aproenca@di.uminho.pt}
}

\institute{High-Assurance Software Laboratory, INESC TEC, Rua Dr. Roberto Frias, 4200-465 Porto, Portugal
          \and
          Dep. Physics, University of Minho, Campus de Gualtar, 4710-057 Braga, Portugal
          \and
          Algoritmi Center, Dep. Informatics, University of Minho, Campus de Gualtar, 4710-057 Braga, Portugal
}

\date{Received: December 6, 2022 / Accepted: March 5, 2023}

\abstract{
HEP-Frame is a new C++ package designed to efficiently perform analyses of data sets from a very large number of events, like those available at the Large Hadron Collider (LHC) at CERN, Geneva.
It mainly targets high performance servers and mini-clusters, and it was designed for natural science researchers with a user-friendly interface to access structured databases.\\
HEP-Frame automatically evaluates the underlying computing resources and builds an adequate code skeleton when creating a data analysis application. 
At run-time, HEP-Frame analyses a sequence of data sets exploring the available parallelism in the code and hardware resources: it concurrently reads inputs from a user-defined data structure and processes them, following the user specific sequence of requirements to select relevant data; it manages the efficient execution of that sequence; and it outputs results in user-defined objects (e.g., ROOT structures), stored together with the used input dataset.\\
This paper shows how a domain expert software development can benefit from HEP-Frame, and how it significantly improved the performance of analyses of large data sets produced in proton-proton collisions at the LHC.
Two case studies are discussed: the associated production of top quarks together with a Higgs boson (\ttH) at the LHC, and a double and single top quark productions at the High-Luminosity phase of the LHC (HL-LHC).
Results show that the HEP-Frame awareness of the analysis code behaviour and structure, and the underlying hardware system, provides powerful and transparent parallelization mechanisms that largely improve the execution time of data analysis applications.
}

\maketitle

\section{Introduction}
\label{introduction}

The new Highly Efficient Pipelined Framework (addressed as HEP-Frame) \cite{HEP-Frame} was developed to efficiently generate, build and execute analysis codes, designed to process large datasets. These can either be acquired by detectors of large experimental collaborations, as in most LHC experiments, or be artificially generated and simulated with the help of Monte Carlo methods. 

In this paper we describe and use the HEP-Frame main features, without loss of generality, in a specific context relevant for the LHC and its High Luminosity phase (HL-LHC), i.e., the development of global analyses of several physics channels, simultaneously. The double and single top quark production channels at the LHC were used as physics signals, in the studies presented in this paper. While the double production of top quarks is usually studied in the semileptonic ($gg\rightarrow t\bar{t}\rightarrow b\ell^+\nu_\ell\bar{b}q\bar{q}'$) and dileptonic ($gg\rightarrow t\bar{t}\rightarrow b\ell^+\nu_{\ell}\bar{b}\ell^-\bar{\nu}_\ell$) decay channels, the single top quark production uses the $t$-channel ($qb\rightarrow q't\rightarrow q'b\ell^+\nu_\ell$) and $Wt$-channel ($gb\rightarrow tW^-\rightarrow b\ell^+\nu_\ell q\bar{q}'$) semileptonic decays. 

Other frameworks have been proposed to re-cast phenomenological analysis  at the LHC~\cite{Conte:2012fm}. 
However, HEP-Frame is the first framework that aims to optimise not only the user interface but also the analysis code performance, taking advantage of the available underlying computing resources during the analyses execution. 
Although HEP-Frame was built to be used in several scientific contexts, we find it particularly useful for challenging applications in High Energy Physics (HEP), given the large amount of data collected by the HEP experiments, like ATLAS~\cite{ATLAS} and CMS~\cite{CMS}. 

The LHC has been colliding beams of protons ($p$) since it started operations in March 2010, at a centre-of-mass energy of 7~TeV. 
Since then, the LHC experiments have been collecting data at increasingly higher centre-of-mass energies, 8 and 13~TeV, until the end of its RUN 2 (at 13~TeV), which ended operation by the end of October 2018. 
A total integrated luminosity of $\sim$150~fb$^{-1}$ was delivered, by the LHC during the RUN 2 alone, to both ATLAS and CMS experiments. 
With $pp$ collisions every 25~ns, the event rate is so significant (40 MHz), that dedicated trigger systems needed to be used, in both experiments, to reduce the event rate of interesting physics, to a manageable level i.e., lower than roughly 1 kHz. 

The importance of the LHC, for the current understanding of the Standard Model (SM) and its fundamental constituents is quite crucial. 
The high production rate of SM particles and the possibility of studying the gauge bosons interactions have allowed to probe the SM with an unprecedented precision, at a new energy scale. 
As a consequence, ATLAS and CMS announced July, 4$^{th}$ 2012, the discovery of a new particle, consistent with the SM Higgs boson with a mass of roughly 125~GeV~\cite{Aad2012tfa,Chatrchyan:2012xdj}. 
This particle was expected to be very rarely produced at the LHC i.e., 1 signal event per tenths of billions of $pp$ collisions. 
Given the level of complexity of the analysis required to identify potential sources of Beyond the Standard Model (BSM) physics, the development of efficient data analysis tools like HEP-Frame, is indeed quite relevant for any research program at the LHC, including phenomenological analysis aiming to propose new strategies to probe the SM. 

In the process of automatically generating an analysis application skeleton and executing its code, HEP-Frame performs, in a consistent and completely transparent way for the user, the following sequential steps:
\begin{itemize}
\item[(1)] automatically builds an analysis code skeleton, adapted to the user input data structure (currently supports ROOT \cite{ROOT} files for LHC case study applications, but users can easily extend this functionality for other file types);
\item[(2)] scrutinises the available hardware resources by looking, not only at the multicore structure of the underlying computing system, but also to the available RAM, computing accelerators (e.g., a GPU), disk space, or other interconnected servers;
\item[(3)] depending on the event size and available computing resources, loads and simultaneously processes several events, taking into account the total available RAM;
\item[(4)] upon user request, can provide different transparent parallelisation of several code operations and,
\item[(5)] delivers results in the form of user defined data structures (ROOT objects like histograms, TTrees, Branches or Leaves, for LHC applications), which may include, not only the input variables judged as relevant for the current analysis, but also new variables needed for later processing, possibly outside of HEP-Frame.
\end{itemize}

Two real-world case studies have been used to provide a quantitative and qualitative assessment of HEP-Frame: the \ttH and top quark analyses.
The former is used to validate the functionality and evaluate the performance improvements of HEP-Frame, for I/O- and compute-bound variations of \ttH.
The latter is used to show how an on-going analysis was developed using this framework.

The rest of this paper is organised as follows. A short overview of the HEP-Frame features is presented in Section~\ref{hep-frame} with a link to a public website with more detailed information, while Section~\ref{analysis} shows how to create a new user analysis. 
Section \ref{casestudy1} presents 3 versions of the \ttH case study to evaluate the performance of HEP-Frame, across different parallelization strategies, in Section \ref{performance}.
The HL-LHC global top quark analysis is explained in Section~\ref{example}, as a second case study of this paper. Section~\ref{conclusions} present our conclusions. 
Appendix I, contains detailed information on how to build and execute an analysis, using an input publicly available ROOT file, which may serve as the basis of any user analysis.

\section{An Overview of the HEP-Frame Tool}
\label{hep-frame}

HEP-Frame is a self contained software tool that builds analysis programs, which efficiently process large sets of data. 
The framework is able to generate codes, across different types of computing platforms (from laptops to clusters, the grid, clouds, etc.), without requiring the user to perform any modification, parallelization, or tuning of the existing code. 
Upon user request, HEP-Frame automatically generates a skeleton of an analysis code in C++, adapted to the user defined input data structure, speeding up the time required to develop a complex event analyses, as required, for instance, at the LHC.

The HEP-Frame package\footnote{The latest version can be downloaded from \url{https://bitbucket.org/ampereira/hep-frame/}.} includes all the necessary programs to successfully build an analysis code. Upon extraction, with\\[-2mm]

{\it unzip <current-hep-frame>.zip},\\[-2mm]

\noindent  the {\it <current-hep-frame>} HEP-Frame main directory is created. 
It contains the directories {\it \textit{lib}}, {\it \textit{scripts}}, {\it \textit{tools}} and  {\it \textit{Analysis}}. 
The latter directory is used to store new analysis applications that can be automatically generated by HEP-Frame. 
To setup the appropriate environment and prepare building the skeleton of the new analysis, the user should move to the HEP-Frame main directory i.e.,\\[-2mm]

{\it cd <current-hep-frame>}\\[-2mm]

\noindent
and compile the code using the HEP-Frame installation script, in the {\it \textit{scripts}} directory. The script can be executed using the shell commands\\[-2mm]

{\it cd scripts}\\
{\it ./install.sh /path-to-the-boost-library/boost}\\[-2mm]

The compilation of the whole code uses, by default, the GNU compiler. HEP-Frame must be linked with two external libraries: BOOST (whose full path should be provided when running the installation script, as shown above) and ROOT (version 5 or 6). Note that, before any analysis code can be generated by HEP-Frame, the user must make sure that HEP-Frame was compiled at least once.

\section{Creating a New User Analysis in HEP-Frame}
\label{analysis}

To create a new analysis, the user must provide an input data file, which is a ROOT file in the case studies presented in this paper, where a TTree with user-defined data structure is expected to be read. 

Following the successful compilation of the whole code, the user is now able to generate the skeleton of a new analysis by running the following shell command (using the scripts in the {\it \textit{scripts}} directory)\\[-2mm]

{\it ./newAnalysis.sh ~<AnalysisName> ~<File> ~<TTree>}\\[-2mm]

\noindent where {\it <{AnalysisName}>} is the name of the new user analysis, {\it {<File>}} is the full path of the input data file and {\it <{TTree}>} is the name of the TTree structure in the user defined ROOT file (where event information is stored) \footnote{The user input ROOT file may also have information stored in the form of ROOT histograms (TH1D, TH2D, and TH3D), which are automatically loaded.}. 

This command creates in the {\it {Analysis}} directory a folder, {\it {AnalysisName}}, which contains the necessary makefiles, and three new folders, {\it {bin}}, {\it {build}} and {\it {src}}. While the first two folders store the generated executable and the required libraries, the {\it {src}} folder stores the automatically generated C++ code skeleton, together with the required data structures. The files contained in the {\it {src}} directory are the ones the user will spend most time on, to adapt the generated code to the specific needs of the analysis. This folder contains the following files:\\[-2mm]

{
 {\it EventInterface.h,} 

 {\it AnalysisName.cxx,} 

 {\it AnalysisName.h,}

 {\it AnalysisName\_Event.cxx,}

 {\it AnalysisName\_Event.h},

 {\it AnalysisName\_cfg.cxx}.\\[-2mm]
}

In the {\it {EventInterface.h}} file, all variables are made automatically available at every selection level (cut) of the analysis (including the ones stored in the input file and any additionally required by the user), upon a successful compilation of the code. No user intervention is expected in this file. 

In {\it {AnalysisName.cxx},} the main analysis class skeleton is implemented (inherited from the DataAnalysis class of HEP-Frame), including the analysis initialisation (once per run), its execution (event by event basis), and finalisation methods (once per run). These are implemented together with the list of cuts that constitute the bulk of the user-defined event selection. Specific global variables of the analysis should be declared in the {\texttt{AnalysisName}} class declaration, in the {\it {AnalysisName.h}} file. These variables should be initialised in the available class constructors of {\it {AnalysisName.cxx}}. 

The user can create as many cuts as necessary. Each cut must return a boolean that indicates if a given event passes the corresponding selection level ({\texttt{bool cutName}}), in the {\it {AnalysisName.cxx}} file. Each cut must be made available to the {\texttt{DataAnalysis}} run-time engine, by calling the method \\[-2mm]

{\small \texttt{anl.addCut("cutName", cutName);}} \\[-2mm]

Note that the user must update the number of cuts variable ({\texttt{number\_of\_cuts}}) in the {\texttt{main}} function of the {\it {AnalysisName.cxx}} file.

The specification of the event information is available in the {\it {AnalysisName\_Event.h}} file, through a C++ class named {\texttt{HEPEvent}}, which is used in the data structure that holds all events in memory. 
The user can add variables to the event, in addition to the ones available in the input ROOT data structure, by declaring them in this file. 
If this is the case, the new variables must be initialised with some default value in the {\it {AnalysisName\_Event.cxx}} file, through the {\texttt{init}} method. 
It should be stressed that the new variables are private to each event and by default they are not automatically saved: the user needs to specify which event variables are relevant to be stored and made available to the user in the output ROOT file (in the {\it {AnalysisName\_cfg.cxx}} file), as explained in Appendix \ref{appendix_1}.

Once the set of user variables is declared, HEP-Frame ensures that the variables will be available at every selection level in the output ROOT data structure. The concept is quite simple: once the user realises the information of a specific variable is relevant at some level of the analysis, HEP-Frame ensures it will be available at any level and, for levels where the information was not yet updated, their default value is used. 

HEP-Frame also supports the creation of auxiliary functions to better organise the code. If these functions need access to partial or full event information, they must receive the argument {\texttt{unsigned this\_event\_counter}} (for internal HEP-Frame management), along with any other user-defined arguments.

The simultaneous analysis of several events requires storing their data in memory. HEP-Frame transparently controls the memory allocation and management, which allows the user to always have access to the event variables available in the {\it {AnalysisName\_Event.h}} and input files as if they were stored in global memory.

\begin{figure}[!htp]
\begin{center}
\includegraphics[height=2.8cm,keepaspectratio]{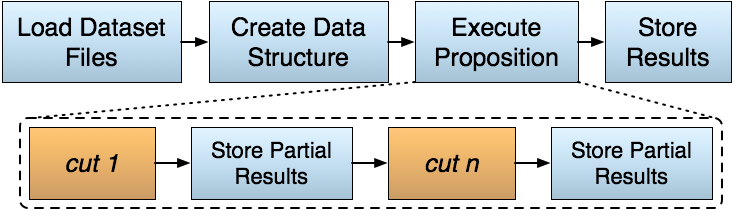}
\caption{Typical event analysis code structure defined by HEP-Frame.}
\label{fig:code_structure}
\end{center}
\end{figure}

The structure of the code required by a HEP-Frame typical event analysis is shown in Figure~\ref{fig:code_structure}. Each element of the analysis chain represents a task that often requires a significant amount of complex C++ code. HEP-Frame implements and dynamically generates the required C++ code for the blue boxes, while the user is responsible to provide the selection criteria for the event analysis (yellow boxes). HEP-Frame automatically handles code generation while scrutinising the underlying computing resources to optimally process the large sets of data.

Following full implementation of the user code in the previous files, and assuming the user has created the new analysis in $\$USER\_WORKDIR$ directory, the code is ready to run on the user defined input file(s). To do so, and to perform a fresh start, a Bash script (e.g. \textit{run\_analysis.sh}) can be used:

{\small
\begin{Verbatim}
# -------------------------------------------
# Dont Forget that if you would like to debug 
# code use ONLY 1 thread alone by performing 
# the following command before running:
# export HEPF_NUM_THREADS=1
# -------------------------------------------
cd $USER_WORKDIR/<AnalysisName>
# Remove old libraries
rm ../../lib/lib/libHEPFrame.a
# Compile and link
make
# run the code
./bin/analysis -f ${inp} -r ${out} -e ${filt}
\end{Verbatim}
}

In the above script, HEP-Frame runs with the input file ({\texttt{\$\{inp\}}}), where the structure (e.g. TTree) used to firstly create the analysis code must exist, creates an output file ({\texttt{\$\{out\}}}) with the user-defined structures and variables to record.
Optionally, it creates a filtered output file ({\texttt{\$\{filt\}}}) with the same structure of the input file, but containing only the events that passed all the user cuts. It should be noted that if the user wants to debug the code, it is advised to use HEP-Frame with a single thread, to avoid receiving concurrent messages from different events that are being processed by several threads at the same time.

\section{The \ttH Case Study}
\label{casestudy1}
%

To validate the HEP-Frame tool and to evaluate its performance simulated data was used that contains signal events from the associated production of top quarks with a Higgs boson ($gg\rightarrow t\bar{t}H$)\footnote{It should be stressed that any analysis in a typical LHC experiment requires large samples of simulated data, in addition to real events, to validate its results against what is expected from the models implemented.}, at the LHC. 
While the top quarks were expected to decay through the leptonic channel ($t\rightarrow bW\rightarrow b\ell\nu_\ell$), the Higgs boson decayed through the dominant decay channel, i.e., $H\rightarrow b\bar{b}$.
The $b$ quarks were detected as jets of particles (labelled $b$-jets) following the hadronization of the initial partons and showering. 
The final state topology of $t\bar{t}H$ events ($gg\rightarrow t\bar{t}H\rightarrow b\ell^+\nu_\ell\bar{b}\ell^-\bar{\nu}_\ell b\bar{b}$), is then characterized by the existence of four jets from the hadronization of $b^(\bar{b})$ quarks, two opposite charged leptons ($\ell^\pm$) produced in the top quark decays, and missing energy from the undetected neutrinos, $\nu_\ell$($\bar{\nu}_\ell$).

The events from $t\bar{t}H$ signal samples were generated at the LHC using the MadGraph5\_aMC@NLO~\cite{Alwall:2014hca} generator. The samples have NLO accuracy in QCD and were generated with the NNPDF2.3~\cite{Ball:2012cx,Ball:2014uwa} parton density functions.
{\scshape MadSpin}~\cite{Artoisenet:2012st} was used to decay the top quarks as well as the heavy bosons ($H$ and $W^\pm$). 
The hadronization, together with the parton shower, was performed by {\scshape Pythia}~\cite{Sjostrand:2006za}. 
All signal events were passed through a fast simulation of a typical LHC experiment (performed by {\scshape Delphes}~\cite{deFavereau:2013fsa}), using the default cards to simulate the ATLAS experiment. One should remark that the theoretical calculation of the $t\bar{t}H$ process has been performed either assuming the Higgs boson has an additional pseudo-scalar (CP) component or within the SM with re-summation precision~\cite{Broggio:2017oyu,Broggio:2019ewu,Kulesza:2018tqz,Broggio:2016lfj,Broggio:2015lya,Maltoni:2015ena}. The study of the CP properties of the $t\bar{t}H$ process through loop corrections has also been done in~\cite{Haber:2022gsn} and attention has been given to the NLO corrections and off-shell effects that impact the observables used to probe the CP nature of the top quark Yukawa coupling~\cite{Hermann:2022vit}. It should also be stressed here that several angular distributions and asymmetries were introduced to study the CP nature of the Higgs boson coupling~\cite{Amor_dos_Santos_2015,Amor_dos_Santos_2017,Azevedo_2020,Azevedo_2021} and interference effects were studied in~\cite{Rui_Santos_2022}.

Although neutrinos cannot be detected directly in \ttH events, their four-momenta may be analytically reconstructed using a kinematic fit. The fit, in addition to imposing energy-momentum conservation to the selected events, assumes the neutrinos come from $W^{\pm}$ boson decays, which, in turn, are originated from the top quark decays, for which mass constraints may be applied~\cite{Santos:2015dja}. 
If sufficient constraints are identified, in a number that exceeds  the number of unknowns, i.e., the 2 neutrino four-momentum, it is possible to fully reconstruct the event kinematics.
This is the case of the \ttH production at the LHC with two opposite charge leptons in the final state.

The code developed for this analysis is a C++ application that includes an event selection with eighteen cuts, organised in a sequential way, as a computational pipeline. 
The measured computation time of each cut significantly varies, from few microseconds to several milliseconds per event, depending on how complex the selection level is.
If the event successfully passes all cuts, than the kinematic reconstruction is applied. 

The kinematic fit aims to reconstruct the undetected neutrinos' four-momenta, as discussed above. 
The reconstruction uses, as constraints, the masses of the top quarks and $W$ bosons, in the following way. The neutrinos from a $W$ decay  must reconstruct, together with one of the charged leptons, the correct $W$ boson mass, fixed to 80.4~GeV. 
This $W$ boson, when paired together with a $b$-jet should, in turn, reconstruct a top quark mass, fixed to 172.5~GeV. 
Once the two isolated leptons are associated with the two reconstructed neutrinos to produce the $W$ bosons, and these are associated to two $b$-jets to reconstruct the top quarks, the kinematic reconstruction attempts to reconstruct the Higgs boson. 
This is done by imposing that two, from the remaining $b$-jets in the event (not associated to the previously reconstructed top quarks), should reconstruct a Higgs boson mass of 125~GeV. 
As there are several possible pairing permutations among the $b$-jets and the charged leptons, a probability is calculated for each permutation, when reconstructing the neutrinos. As the system of constraint equations have quadratic forms~\cite{Santos:2015dja}, if a solution is possible, normally there are several available\footnote{In many events, following the DELPHES simulation of ATLAS, the kinematic properties of the detected jets and leptons are so distorted due to resolution effects, that it is quite impossible to make these events compatible with a true \ttH event. In this case no solution will be possible, compatible with the mass constraint equations.}. If there are solutions, the one with highest probability is considered to be the correct solution, with the correct combination of particles. The last cut of the event selection is precisely this kinematic reconstruction, which discards the event if no solution was found.


The performance of reconstructions that attempt to compensate for detector resolutions, or any other effects that have an impact on the kinematic properties of the events, is too low, making these extensive analyses unfeasible. If several solutions are to be tested per event, and if there are millions of events to be analysed, rapidly the analysis performance becomes very much limited by the available computing power. Unfortunately, it is very common that the user never cares about testing the system, during the process of building and running an analysis. 
HEP-Frame is particularly useful in this context: it adapts to the underlying computational resources of the server and the characteristics of the analysis during its execution. This allows HEP-Frame to adapt to the server and analysis without any previous knowledge and no interaction of the user.

To test the HEP-Frame performance for the \ttH case study, three versions of the dileptonic \ttH analysis were considered:

\begin{description}
    \item[\texttt{ttH\_as} (\textit{\underline{a}ccurate detector \underline{s}ystem}):] this version assumes  accurate resolution detectors and the DELPHES simulation of the ATLAS response is taken exactly as is, the data measured by the ATLAS detector is considered 100\% accurate when reconstructing the event. This behaves as a I/O-bound code in most compute servers.
    
    \item[\texttt{ttH\_sci} ( \textit{detector \underline{s}ystem with a \underline{c}onfidence \underline{i}nterval} ):] this version assumes a $1\%$ random uncertainty, associated to the ATLAS detector energy and momentum measurements, due to resolution effects. This defines a confidence interval\footnote{This interval should not be misinterpreted as the usual statistical confidence interval. Here the measurements are sampled within the specified uncertainty for the possible measurements, on an event by event basis.}. An extensive sampling of events was performed, where the particles measured energies and momenta were varied within the fixed uncertainty during the reconstruction. Only the highest probable solution was considered, as explained above.
    This version recreates 1024 pseudo experiment samples of the original one, where each requires the generation of 30 different pseudo-random numbers (PRNs), to a total of 30 Ki numbers per event, leading to a compute-bound code.
    
    \item[\texttt{ttH\_scinp} (\texttt{sci} \textit{with a \underline{n}ew \underline{p}ipeline}):] two cuts of the event selection were replaced to perform different operations on the data elements, maintaining the same overall cut dependencies.  For this version, only 128 pseudo experiment samples were recreated, within the same confidence interval of the measurements performed for the previous version of the code (\texttt{ttH\_sci}).
    This version is also compute-bound, but is less compute intensive than \texttt{ttH\_sci}.
\end{description}

Hep-Frame was already used when preparing the \textit{“Report from Working Group 1: Standard Model Physics at the HL-LHC and HE-LHC”} for the European Strategy for High Energy Physics \cite{HE-LHC}. In these studies, datasets produced for the HL-LHC of the ATLAS Experiment were analysed for all Standard Model physics processes. As expected, no major difficulties were observed on the analyses developed with HEP-Frame even when running and submitting the jobs to the CERN clusters.

\section{The HEP-Frame Performance with the \ttH Case Study}
\label{performance} 

Four different compute servers were selected for the quantitative evaluation of the HEP-Frame performance:

\begin{itemize}
    \item a dual-socket server with 12-core Intel Xeon E5-2695v2 Ivy Bridge (IB) devices (@2.4 GHz nominal, with 64 GiB RAM), coupled with a NVidia Tesla K20 (2496 CUDA cores and 5 GiB of GDDR5 memory).
    
    \item a dual-socket server with 16-core Intel Xeon E5-2683v4 Broadwell (BW) devices (@2.1 GHz nominal, 1.7 GHz nominal with AVX2, with 256 GiB RAM).
    
    \item a dual-socket server with 24-core Intel Xeon Platinum 8160 Skylake devices (@2.1 GHz nominal, 1.4 GHz nominal with AVX-512, with 192 GiB RAM).
    
    \item a single-socket server with 64-core Intel Xeon Phi 7210 device, KNL (@1.3 GHz nominal, 1.1 GHz nominal with AVX-512, 4-way simultaneous multithreading, with 16 GiB of embedded HBRAM and 192 GiB of RAM).
\end{itemize}

The performance of the different versions of the \ttH analysis code, implemented in HEP-Frame, is briefly discussed in the next subsections. This evaluation focuses on:

\begin{itemize} 
\item automatic and transparent tuning of HEP-Frame to the underlying computing resources and automatic multithread parallelisation of the code (Section~\ref{multithreading}); 
\item HEP-Frame advanced task parallelisation of the code (Section~\ref{parallelisation}); 
\item offloading computationally heavy tasks to the GPU accelerator (Section~\ref{offloadGPU}).
\item the overall performance of HEP-Frame with additional optimisations (Section~\ref{parallelisation2}).
\end{itemize}

A more detailed description of the individual parallelization strategies, from the computing point of view, can be found in the literature \cite{HEP-Frame,AMPPHD,SchedulerPaper,HEPF,ICMA}.


\subsection{Hardware Aware Multi-Threading}
\label{multithreading}

HEP-Frame implements multiple strategies to parallelize the execution of an event analysis application, each with a specific purpose.
The main focus of this framework is to efficiently use the computational resources available in heterogeneous servers, i.e., multicore and manycore CPU devices coupled with manycore and GPU accelerators, so that applications process more data in less time.
All optimisations employed by HEP-Frame are completely transparent to the user, not requiring any knowledge of the computing platform nor any interaction from the user.

The main parallelization approach in HEP-Frame relies on distributing the workload among a pool of automatically created and managed threads.
This workload can either be related to input file reading with event pre-processing, or associated to full event processing, level by level. While the former depends on the complexity of the input ROOT data structures, the latter is determined by the algorithms used in each selection level, and both may be quite compute intensive.

To parallelize this mixed workload, HEP-Frame dynamically manages during run-time the amount of threads assigned to input reading and to event processing, adapting to its requirements without any user interaction. When HEP-Frame receives a batch of input files it also automatically manages the file pre-processing and the generation of an intermediate output file (during execution) and the final user defined data (when the application finishes execution).

\begin{figure}[!htp]
	\begin{center}
	\includegraphics[width=10cm,keepaspectratio]{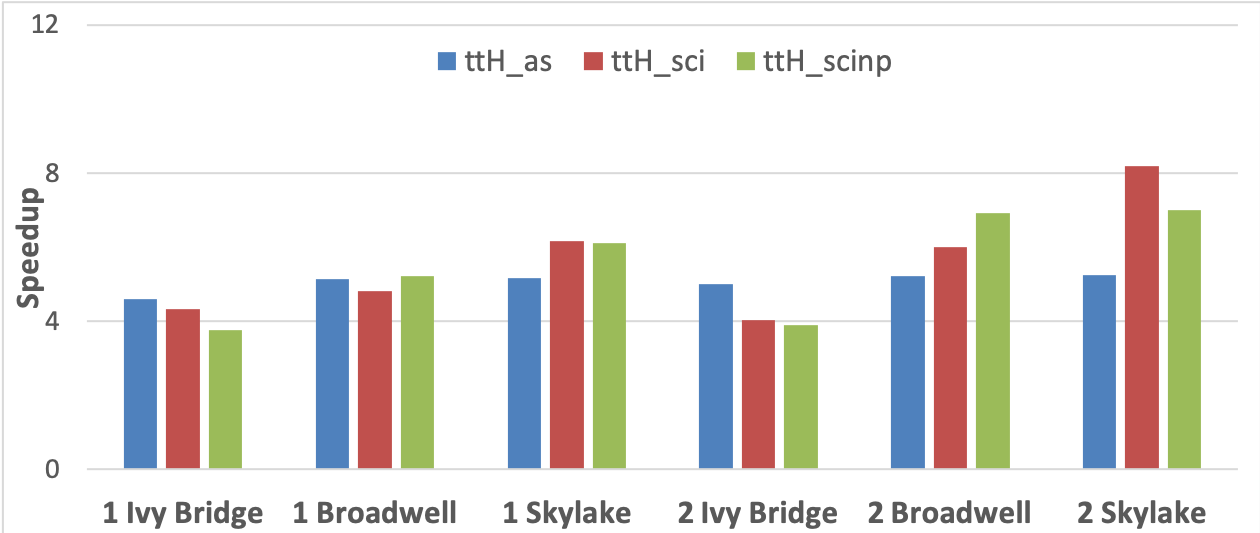}
		\caption{Speedup of the parallel \ttH analyses with HEP-Frame multithread parallelisation \textit{vs} a standard multiprocess parallelisation for the same number of processes/threads on a server with single or dual multicore devices.}
		\label{fig:original_speedup_mt}
	\end{center}
\end{figure}

The framework has a \textit{simple} and an \textit{advanced} strategy to parallelize the event processing.
The \textit{simple} strategy assigns different events to different computing threads, each processing the whole pipeline of cuts, individually for each event. 
The workload is dynamically distributed among the processing threads due to the irregularity of the pipeline processing. Events that pass fewer cuts will result in faster execution pipeline than events that are processed by the whole pipeline, causing unpredictable irregularity in the workload.
The performance of this strategy is compared against a standard multiprocess approach, where the user has to launch each individual process with different input files and ensure that they successfully finish executing, in Figure~\ref{fig:original_speedup_mt}.
The \textit{advanced} strategy is discussed in section \ref{parallelisation}, below.

HEP-Frame outperforms the standard multiprocess parallelization, for the same amount of threads/processes, in all variations of the \ttH analysis, with improvements up to 5.2x, 8.2x, and 7.0x for the \texttt{ttH\_as}, \texttt{ttH\_sci}, and \texttt{ttH\_scinp}, respectively. 
These performance improvements were achieved by a combination of the simultaneous data reading and processing of the analyses and adequate distribution of the irregular workload among the computing threads (this assessment is further detailed in \cite{HEP-Frame}).
Such improvements are not restricted to a single server, but are consistent among the single-socket and dual-socket servers of different architectures (Ivy Bridge, Broadwell, and Skylake), as HEP-Frame automatically adapts in run-time the amount of threads and parallelization strategies to the characteristics of the server.

\subsection{Pipeline Aware Cut Parallelisation}
\label{parallelisation}

The initial execution order of the cuts in the processing pipe-line is defined by the user, following a simple logical reasoning determined, in most cases, by the characteristics of signal events. Although this is normally the typical procedure, when first defining the event selection, this may not be the best ordering, in terms of computational efficiency. This is particularly relevant for large datasets and, the ones collected at the LHC, are good examples of those.

Reordering the cuts, while respecting the dependencies among cuts to ensure the correctness of the results, often leads to a faster execution of the pipeline.
If the cuts that discard more data elements, are placed earlier in the pipeline, and the heavier cuts in later stages, fewer data will be processed by the heavier, computational intensive, cuts, reducing the overall execution time of the pipeline.
This reordering must take into account the amount of events that each cut filters out, as well as their execution time.

The order of the cuts in the pipeline has a significant impact on performance, since having the most compute intensive cuts at the end of the pipeline is more efficient as they are applied to fewer events than if they were placed in the beginning.
Static ordering of the cuts is not a recommended approach for these applications, since the behaviour of the cuts cannot be measured before executing the application, and may even change during its execution.
Alternatively, HEP-Frame dynamically optimises the ordering of the cuts during execution, by distributing them among the available processing threads. This is done not only for the cuts in the same event but also when multiple events are simultaneously processed.

Parallelizing the workload at the cut level ensures that a more efficient load balance can be obtained for both memory- and compute-bound applications due to the smaller size of each individual task, as opposed to parallelize only at the event level. Traditional list schedulers are extremely efficient at managing pipelines of tasks that do not filter out events, but are not designed to schedule cuts. Their lack of support for task-level parallelism of a single or multiple events does not ensure the most efficient balance of data and propositions among threads. This may lead to the unnecessary execution of computationally intensive cuts. HEP-Frame's novel strategy of simultaneously processing cuts of the same and different events, while reordering and respecting cut dependencies, reduces the computational load and leads to a faster adaptation to irregular workloads, thus reducing the overall execution time compared to alternative approaches.

\begin{figure}[!htp]
	\begin{center}
	\includegraphics[width=10cm,keepaspectratio]{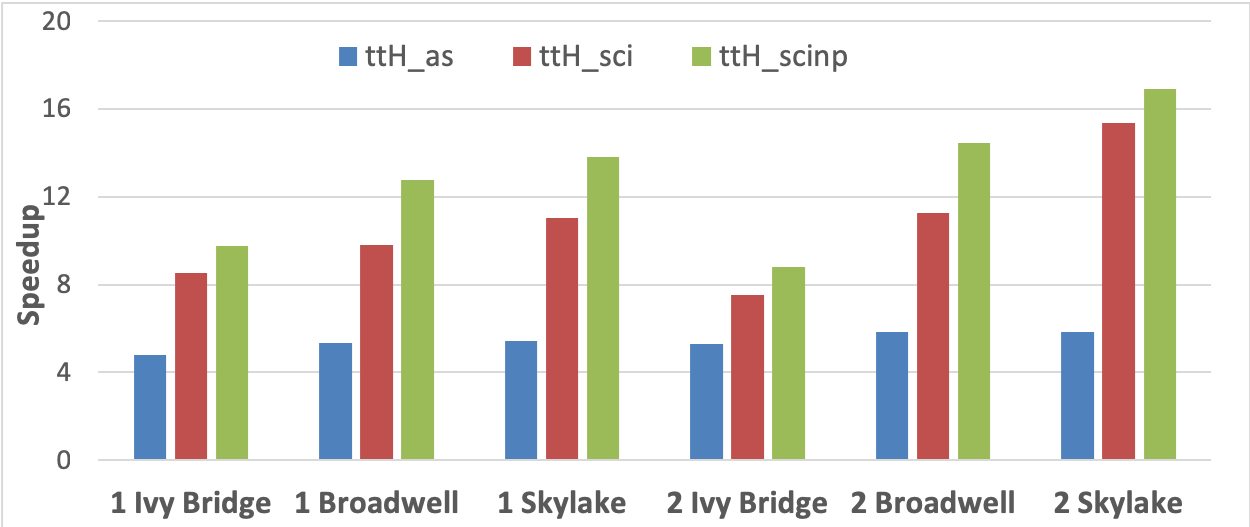}
		\caption{Speedup of the parallel \ttH analyses with HEP-Frame task parallelisation \textit{vs} a standard multiprocess parallelisation for the same number of processes/threads on a server with single or dual multicore devices.}
		\label{fig:original_speedup}
	\end{center}
\end{figure}

The performance of the multithreaded \ttH analyses, using a multiprocess approach, was compared against their implementations in HEP-Frame using one and two Xeon devices of the Ivy Bridge, Broadwell and Skylake micro-architectures, as shown in  Figure~\ref{fig:original_speedup}.
Both parallelisations use a single thread per physical core on the server, as preliminary tests showed that using hardware support for simultaneous multithreading in each core (addressed as Hyper-Threading by Intel) did not provide noticeable performance improvements. Note that the initial order of the cuts in the pipeline was defined by the physicist responsible for the case study and it was considered the performance and filtering ratios of the cuts in this initial order, which were obtained through an \textit{ad-hoc} analysis of the software. Worse initial organisation of the cuts could have been used to provide larger performance improvements, but they would not be indicative of a real case study.
HEP-Frame significantly improved the performance of all multithreaded implementations:

\begin{description}
    \item[\texttt{ttH\_as}:] up to 6x faster, mostly due to simultaneous event reading and processing, which mitigates I/O bottlenecks.
    
    \item[\texttt{ttH\_sci} and \texttt{ttH\_scinp}:] 15x and 17x speedups, respectively, mostly due to the pipeline reordering scheduler.
    
    \item[\texttt{ttH\_scinp}:] performance improvements also due to a worse initial pipeline order than \texttt{ttH\_sci}.
\end{description}

The performance gap between HEP-Frame and multiprocess increases proportionally to the number of cores in the server, as shown by the improved speedup when using dual Broadwell and Skylake devices over a single device.
This proves that the multiprocess approach efficiency diminishes opposed to HEP-Frame, especially when dealing with a high number of workers, and that this difference is not only related to the pipeline reordering.

Finally, the whole event processing, from single or multiple input files, is transparently managed by HEP-Frame, which avoids the overhead of an user close control of the execution of multiple processes. An in-depth analysis of the lower-level computational behaviour of HEP-Frame, addressing how this framework handles the specifics of I/O, memory, and computing bottlenecks of these applications, is available in \cite{HEP-Frame}.

It should also be mentioned that the initial ordering of the pipeline in these \ttH analyses, as defined by domain experts, had luckily most of their cuts filtering out more events at the beginning of the event selection, leaving the heavier cuts to the final pipeline stages.
Applications with worse default pipeline orders would benefit even more from the HEP-Frame pipeline reordering.

\subsection{Offloading Computationally Heavy Tasks to Accelerators}
\label{offloadGPU}

Porting code from external libraries to a GPU device (or other accelerator device) is not always possible or feasible in the short available time of a domain expert. 
In these cases, HEP-Frame can automatically take advantage of these computing accelerators by offloading computationally heavy tasks to them, freeing the host multicore devices to process the remaining parts of the applications.
Since cuts cannot be offloaded, as it would require the user to provide the GPU code, which is often not possible due dependencies on external libraries, such as ROOT, these devices should be used to accelerate tasks common among various analyses.
HEP-Frame has already one of these heavy and highly used tasks implemented to work in the offload mode: an efficient pseudo-random number generator (PRNG) for large datasets \cite{ICMA}.

In the execution time measurements with the \ttH case study, HEP-Frame took advantage of the available GPU device in the IB server: it used the Mersenne Twister PRNG, the default PRNG provided by ROOT, implemented in MKL \cite{MKL} for the multicore-only servers and implemented in cuRAND \cite{curand} when offloading to a GPU device.
The Mersenne Twister PRNG implementation in ROOT is considerably slower than these two alternatives.
The use of the Kepler GPU improved the performance of the \texttt{ttH\_sci} and \texttt{ttH\_scinp} by $70$x and $12$x, respectively, compared to generating PRNs using ROOT; this improvement depends on the number of PRNs required by each one of these applications. Since PRN generation is offloaded to the GPU, HEP-Frame can take advantage of the additional CPU resources to process the analysis cuts.
While a study of the computational performance of using GPUs with HEP-Frame is out of the scope of this communication, a deeper analysis of the impact of this approach can be found in \cite{PRNG-Broker}.

\subsection{Overall Performance}
\label{parallelisation2}

\begin{figure}[!htp]
	\begin{center}
	\includegraphics[width=9cm,keepaspectratio]{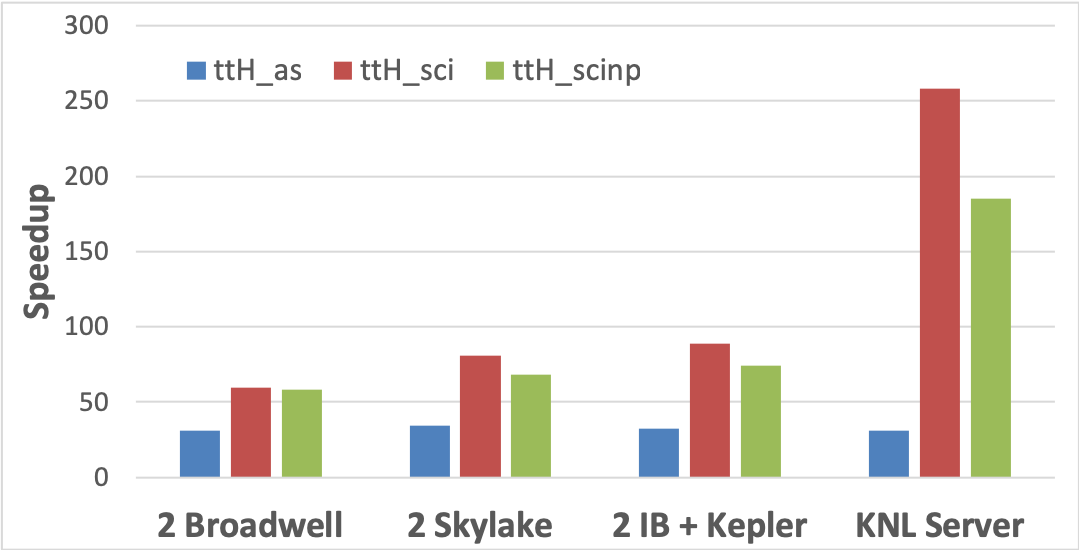}
		\caption{Overall speedup of the \ttH analysis on HEP-Frame \textit{vs} their original sequential implementations.}
		\label{fig:vsseq}
	\end{center}
\end{figure}

Figure~\ref{fig:vsseq} compares the overall performance of the three versions of the \ttH analyses, implemented with HEP-Frame, with their original sequential implementation.
The execution times were measured on the best multicore servers (with dual Broadwell and Skylake devices), on the Ivy Bridge server with a Nvidia Kepler GPU, and on the Intel KNL server.

HEP-Frame provided a significant execution time improvement for every case study version, confirming its performance portability across multiple platforms with significant architectural differences.
It adapted well to irregular compute-bound code, with speedups up to 252x and 185x for the \texttt{ttH\_sci} and \texttt{ttH\_scinp} versions, respectively, on the KNL server.
It also efficiently handled the I/O-bound \texttt{ttH\_as} version, with a speedup 30x for every server, due to its dynamic tuning of the amount of threads assigned for simultaneous input reading and event processing.\\[-1mm]

The KNL server outperformed every other server mainly due to its core count and greater vectorization capabilities: two AVX-512 vector units per core, while Skylake has only one AVX-512 vector unit and Broadwell and Skylake have AVX units to operate on 256 bits. 
These two also suffered significant down clock frequency due to the AVX instructions, which is less severe on the KNL.

\section{ The Top Quarks Analysis Case Study}
\label{example}

This section presents the second case study, the analysis of top quarks at the HL-LHC.
This case study was implemented using HEP-Frame, which was also used to manage its efficient execution and produce the results presented in this section, using large sets of simulated data through Monte-Carlo methods.
The choice of performing a global analysis of double and single top quark production at the HL-LHC simultaneously poses concrete challenges, which HEP-Frame can easily address.

The pipelines of the event selections presented here, follow closely the ones made available by the ATLAS Collaboration, for the double top quark production~\cite{Aaboud:2018uzf,Aaboud:2016bit}, and single top quark search~\cite{Aaboud:2016ymp,Aaboud:2017aqp,Aaboud:2016lpj}, at the LHC.
As discussed in Section~\ref{introduction}, while the double production of top quarks concentrates on both the semileptonic ($gg\rightarrow t\bar{t}\rightarrow b\ell^+\nu_\ell\bar{b}q\bar{q}'$) and  dileptonic ($gg\rightarrow t\bar{t}\rightarrow b\ell^+\nu_{\ell}\bar{b}\ell^-\bar{\nu}_\ell$) decay channels, the single top quark production uses the $t$-channel ($qb\rightarrow q't\rightarrow q'b\ell^+\nu_\ell$) and $Wt$-channel ($gb\rightarrow tW^-\rightarrow b\ell^+\nu_\ell q\bar{q}'$) semileptonic decays, alone. 

\subsection{Event Selection at the HL-LHC}
\label{selection}

The pipeline of cuts used to define the global event selection of top quark events at the HL-LHC aims to efficiently identify signal regions, corresponding to the different physics channels under study, to be, as much as possible, free of SM backgrounds. The events that pass all cuts are used to build specific angular distributions that are sensitive to BSM. As these angular distributions involve the knowledge of the four-momentum of top quarks, full reconstruction of the kinematic properties of final state particles is mandatory, in particular for the undetected neutrinos.

The signal regions, targeted by the event selection, were divided into:

\begin{itemize}
\item[(1)] three semileptonic final states, corresponding to the production of $\ttbar$ and single top quark events through the $t$- and $Wt$-channels, where exactly 1 isolated ($\Delta R\footnote{The pseudorapidity $\eta$ of a particle is defined as $\eta$ = − ln[tan($\theta$/2)], where $\theta$ is the particle's polar angle. The pseudorapidity difference between two particles ($\Delta\eta$) together with their azimuthal angle difference $\Delta\phi$ are used to defined the $\Delta R=\sqrt{(\Delta\eta)^2+(\Delta\phi)^2}$ distance.} < 0.4$) lepton ($e^{\pm}$ or $\mu^{\pm}$) is found, and

\item[(2)] two dileptonic final state topologies, from $\ttbar$ and $Wt$ single top quark associated production, where exactly 2 isolated and opposite sign charged leptons ($e^{\mp}\mu^{\pm}$) of different flavours, are present.
\end{itemize}

A cut on the missing transverse energy ($E_T^{miss}$) was also applied to events, $E_T^{miss}$>30~GeV.  Events were further classified according to the number of jets existing in three non-overlapping $\eta$ regions corresponding to,
\begin{itemize} 
  \item one central region, where the jets $\eta$ obey the condition $|\eta|$<2.5 (labelled as Region I),
  \item a region where the jets $\eta$ was in the range 2.5 < $|\eta|$ < 2.75 (Region II), and
  \item a forward region, where 2.75 < $|\eta|$ < 3.50 (Region III).
\end{itemize}

These specific signal regions were defined following $\eta$ acceptance regions commonly used in $\ttbar$ and single top quark event selections (for both semileptonic and dileptonic final states) ~\cite{Aaboud:2018uzf,Aaboud:2016bit,Aaboud:2016ymp,Aaboud:2017aqp,Aaboud:2016lpj}. 
The events were split among different signal bins, according to lepton, jets and $b$-jets multiplicities:

\begin{itemize}
\item[(1)]signal bins from $\ttbar$ semileptonic (dileptonic) decays, were populated with events with exactly 1 charged lepton (2 opposite sign charged leptons), at least 4 jets in Region I and exactly 2 $b$-jets (at least 2 jets in Region I and exactly 2 $b$-jets). No jets in Regions II and III were allowed in $\ttbar$ signal bins. For the $\ttbar$ dileptonic decays, the invariant mass of the two leptons ($M_{l^+l^-}$) was required to be above 40~GeV; 

\item[(2)]bins of single top quark $t$-channel events, were populated, if they had exactly 1 charged lepton, 1 jet in Region I and 1 jet in either Region II or III, with exactly 1 $b$-jet; 

\item[(3)]bins from signal events of $Wt$ single top quark production, which decayed through the semileptonic (dileptonic) channel were filled with events with exactly 1 charged lepton (2 opposite sign charged leptons), 3 jets in Region I and exactly 1 $b$-jet (1 or 2 jets in Region I and exactly 1 $b$-jet). No jets in Regions II or III were allowed in the events. Moreover, for the semileptonic channel, in order to reduce the $\ttbar$ background, a cut on the $W$-boson transverse mass ($M^T_W$) was applied above 50~GeV.
\end{itemize}

As explained in Section~\ref{analysis}, the cuts discussed above were implemented in the $AnalysisName.cxx$ file. As an example we show how to do it for the $E^{miss}_T$ cut. As its information is relevant for later use, we also show how to declare a new variable ($ETmis$) to store its value and save it to the output ROOT file, after initialisation. We start by declaring the variable in $AnalysisName\_Event.h$,\\[-2mm]

{\small
\begin{Verbatim}
class HEPEvent {
    TTree *fChain;
public:               
    ...            
    // Add your own event variables here!
    double ETmis; 
    ...         
} 
\end{Verbatim}
}

\noindent 
Next, we initialise $ETmis$ in $AnalysisName\_Event.cxx$,
{\small
\begin{Verbatim}
bool HEPEvent::init (void) {
    // Initialize all my variables of this event 
    ...           
    ETmis = -9999.;
    ...            
}         
\end{Verbatim}
}

\noindent 
The $ETmis$ variable can now be added to the user list of variables to be recorded, in $AnalysisName\_cfg.cxx$, right after the end of the $writeVariables$ method. This will make the variable available in the output ROOT file, for later use. 
 Following the previous example, the lines of code required to add $ETmis$ to the relevant list of variables are,

{\small
\begin{Verbatim}
void AnalysisName::writeVariables (void) {
}                          
// Write here the variables and expressions
// to record per cut
#ifdef RecordVariables     
ETmis  
#endif
\end{Verbatim}
}     

The variables considered as relevant, which are indicated to HEP-Frame as shown above, will be automatically stored in a TTree, one per cut, and stored in a ROOT file at the end of the analysis execution.
Once the $ETmis$ variable has been declared, it can be used in a cut, in the $AnalysisName.cxx$ file. The cut must return a boolean, which indicates if a given event passes the cut, and it is advised that the user creates a {\it cut\_evaluation} function (just to better organise the code, not done automatically by HEP-Frame) to perform all the user requested actions upon a true return of the cut result. The lines of code resemble to,

{\small
\begin{Verbatim} 
void cut_evaluation (unsigned this_event_counter) {
    // A cut can evaluated variables
    ETmis *= 1.05;  // scale factor to ETmis
}

bool cut (unsigned this_event_counter) {
    // A cut should return true or false
    if  ( met_met <= 30000. ) {
        return false;
    } else {
        ETmis  = met_met;
        cut_evaluation(this_event_counter);
        return  true;
    }
}
\end{Verbatim}
}

\begin{figure}[!htp]
\begin{center}
\includegraphics[height=6cm,keepaspectratio]{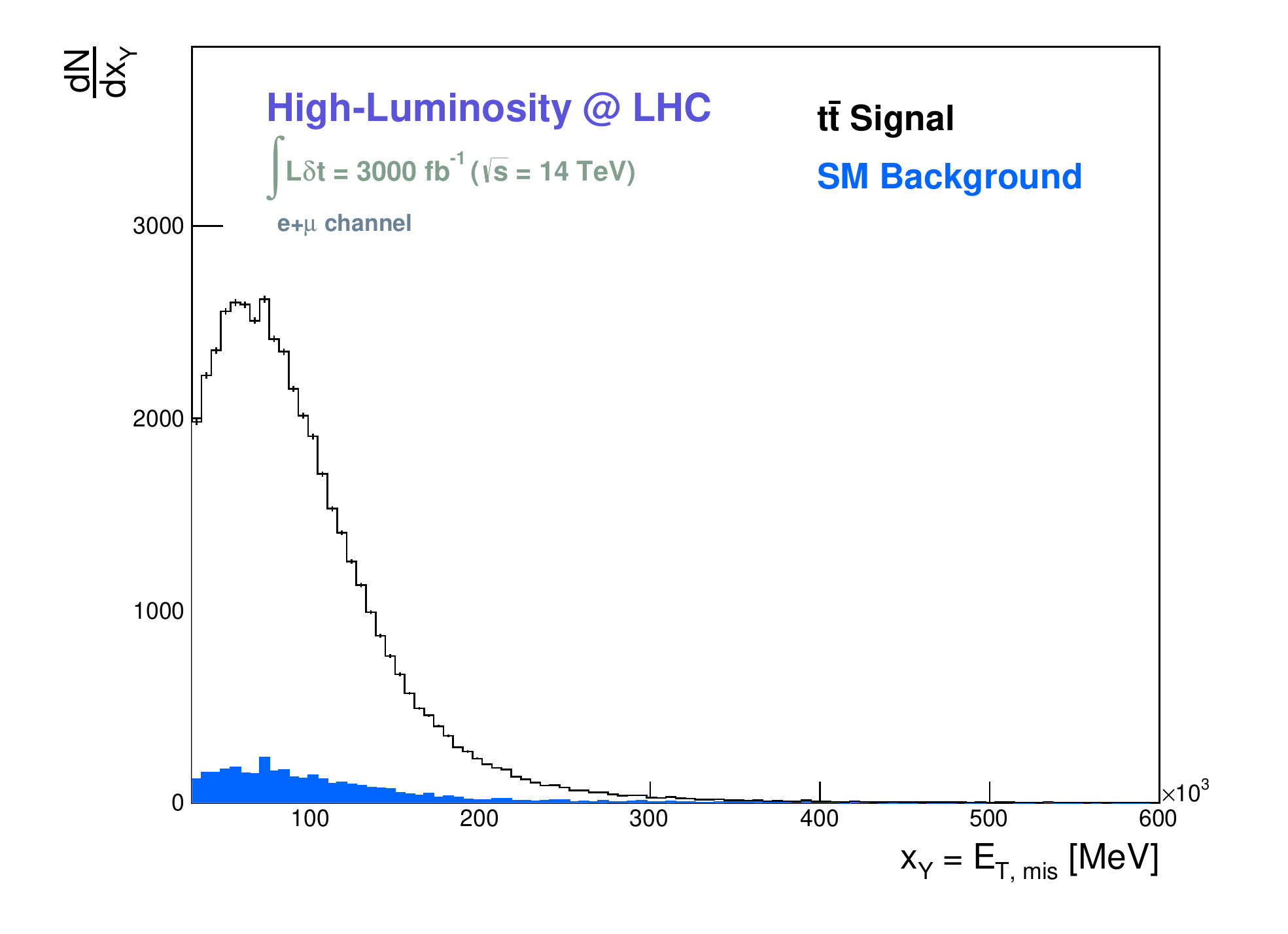}
\includegraphics[height=6cm,keepaspectratio]{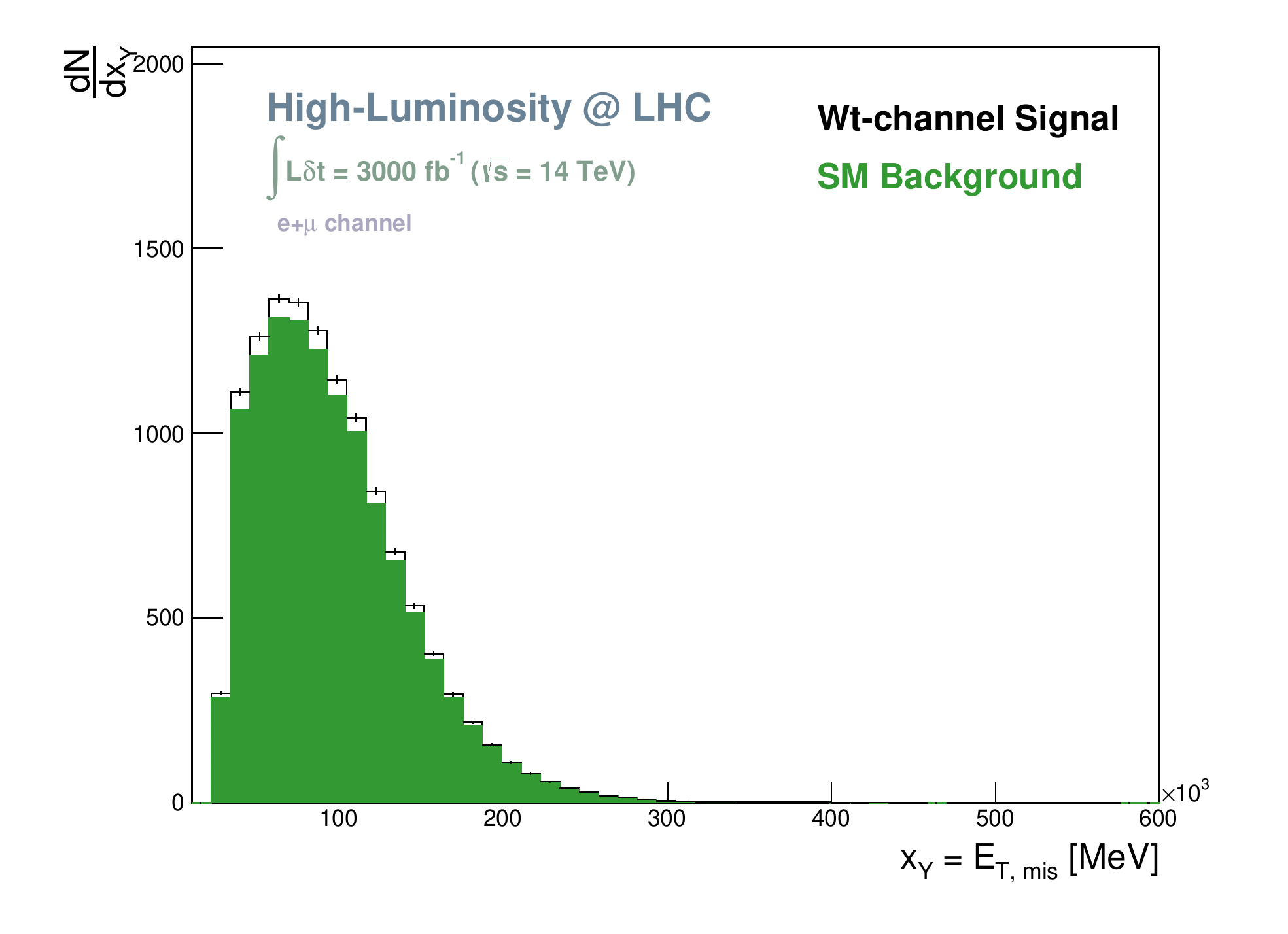}
\caption{The missing transverse energy for $t\bar{t}$ semileptonic events (left) and single-top quark $Wt$ dileptonic events (right).}
\label{fig:etmis}
\end{center}
\end{figure}

To call the cut in the same $AnalysisName.cxx$ file, we just need to include the following lines of code, in the main method, not forgetting to increment the counter of cuts,

{\small
\begin{Verbatim}
int main (int argc, char *argv[]) { 
    ...           
    // Do not forget to set the number of
    // cuts to add!!!!! 
    // You must also set them on the Makefile 
    unsigned number_of_cuts = 1; 
    ...         
    // Add the cuts using the addCut method 
    anl.addCut("cut", cut);
    ...         
    return 0;
}
\end{Verbatim}
}
In Figure~\ref{fig:etmis} we show the missing transverse energy for $t\bar{t}$ semileptonic events (left) and single-top quark $Wt$ dileptonic events (right), after full event selection. The signal and all SM backgrounds are shown for completeness, assuming the full luminosity at the HL-LHC (3000~fb$^{-1}$).

\subsection{Kinematic Reconstruction of Signal Events}
\label{topquark_rec}

Following the event selection, full kinematic reconstruction is applied to each signal region i.e., to the semileptonic and dileptonic final states from $\ttbar$, single top quark $t$-channel and $Wt$-channel decays.  In each signal region and for each possible jet and lepton combination, a $\chi^2$ function was minimised to derive the  four-momentum of the undetected neutrino(s) and reconstruct the top-quark(s) and $W$-boson(s) masses. The solution, among all possible combinations of jets and leptons, which minimises the value of the $\chi^2$, defined as,

\begin{equation}
	\chi^2  =   \sum_{k=1(2)} \frac{\left(m^{\mathrm{reco}}_{j \ell \nu}-m_{t}\right)^2}{\sigma_{t}^2}
	             +  \sum_{m=1(2)} \frac{\left(m^{\mathrm{reco}}_{\ell \nu}-m_W\right)^2}{\sigma_W^2},
	\label{eq:chi2}
\end{equation}

\noindent is chosen. The indices $k$ and $m$ can either take the value 1 or 2, depending on the number of top quarks and $W$-bosons expected in the events. While for $\ttbar$ events 2 top quarks and 2 $W$-bosons should be present (in both semileptonic and dileptonic final states), in single top quark events from $Wt$ associated production, only 1 top quark and 2 $W$-bosons should be reconstructed, while for the $t$-channel only 1, of each, should exist. In the $\chi^2$ definition, $m^{\mathrm{reco}}_{j \ell \nu}$ ($m^{\mathrm{reco}}_{\ell \nu}$) represents the reconstructed invariant mass of the top quark ($W$-boson), for the particular combination of jets and leptons under consideration.  

\begin{figure}[!htp]
\begin{center}
\includegraphics[height=6cm,keepaspectratio]{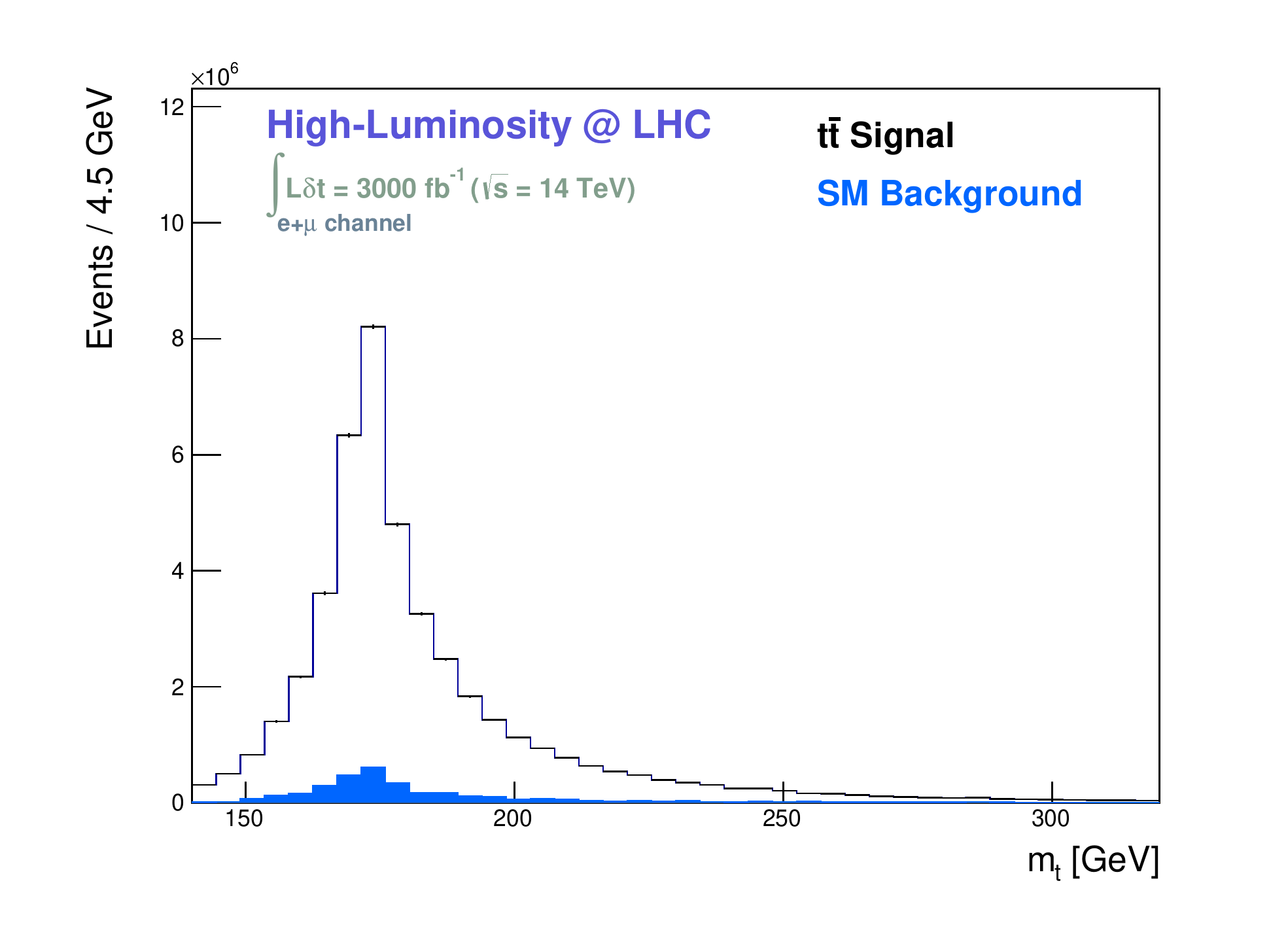}
\includegraphics[height=6cm,keepaspectratio]{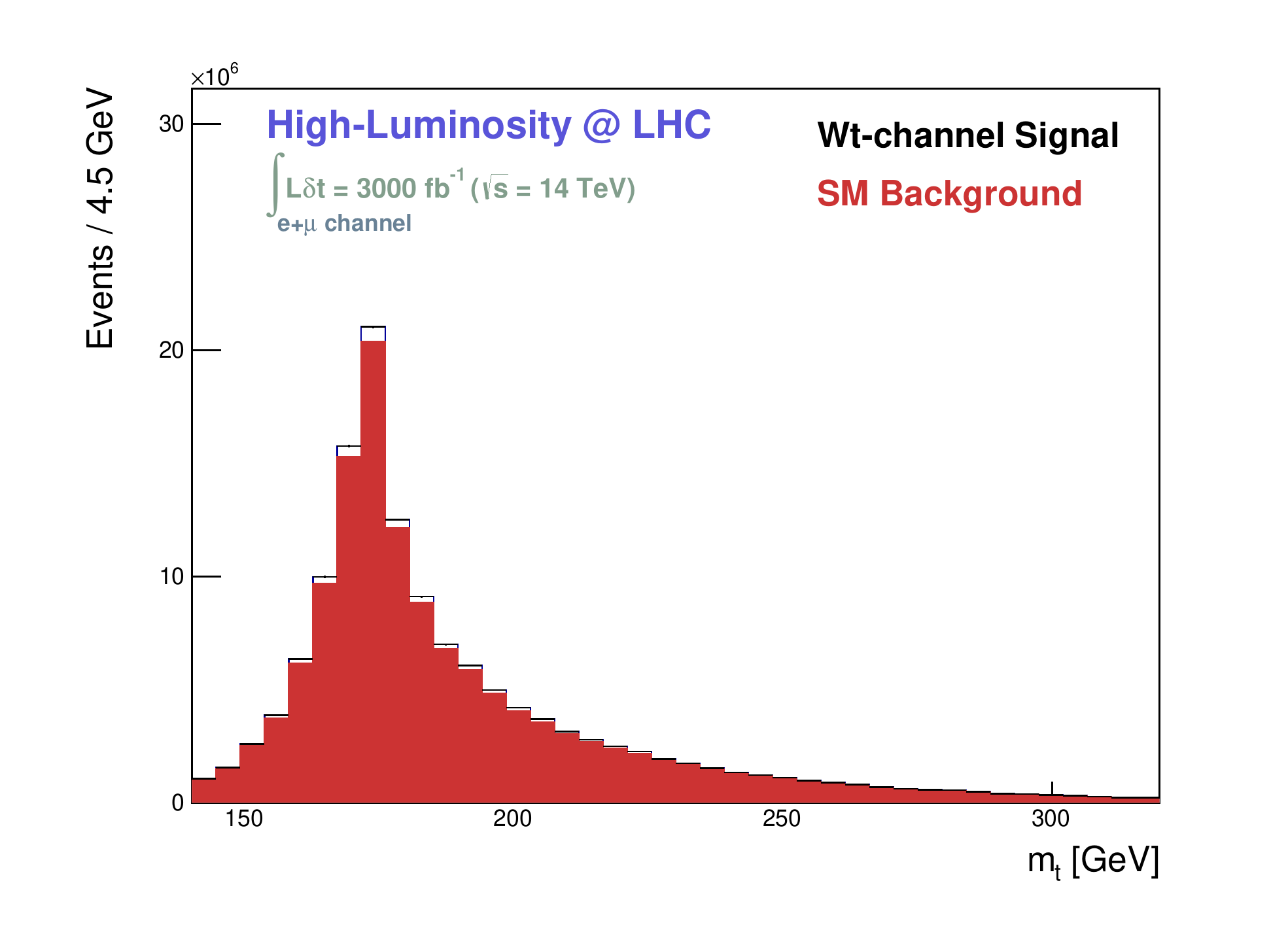}
\caption{The reconstructed top quark mass in $t\bar{t}$ semileptonic events (left) and single-top quark $Wt$ events (right).}
\label{fig:top_rec}
\end{center}
\end{figure}
The central values of the top quark mass ($m_t$) and $W$-boson mass ($m_W$), were fixed to 172.5~GeV and 80.4~GeV, respectively. The corresponding widths, $\sigma_t$ and $\sigma_W$, were set to 11.5~GeV and 7.5~GeV, respectively. In the minimisation procedure, the $E^{miss}_T$ is assumed to be the transverse momentum of the undetected neutrino(s). While for the semileptonic final states only the $p_Z$ component of the neutrino four-momentum remains to be determined, for the dileptonic final states, two neutrinos 
must be fully reconstructed. This implies, splitting the  $E^{miss}_T$ between the two neutrinos and determine, using the mass constraints from the $\chi^2$ function, their $p_{Z}$ component.
In Figure~\ref{fig:top_rec} we show the top quark mass reconstruction for semileptonic  $t\bar{t}$ events (left) and single-top quark $Wt$ events (right), after event selection. The signal and all SM backgrounds are shown for completeness, assuming the full luminosity at the HL-LHC (3000~fb$^{-1}$).

\section{Conclusions}
\label{conclusions}

In this paper we presented the Highly Efficient Pipelined Framework (HEP-Frame), a C++ tool created with two purposes: (i) to help the development of analysis and simulation codes that process large amounts of data; (ii) to ensure the efficient usage of the underlying parallel hardware available in servers with and without computing accelerators, namely GPU devices.
HEP-Frame provides an easy user interface to develop code, by automatically creating analysis skeletons based on input data structures, significantly speeding up code development.
It implements several parallelization strategies, so that the analysis code dynamically and transparently adapts to the available hardware, without the need of any user intervention, to ensure efficient execution of the applications across a variety of systems.

Two examples of High Energy Physics analysis, at the LHC, were discussed in this paper, as case studies:  the associated production of top quarks together with a Higgs boson (\ttH), and the double and single top quark production at the HL-LHC.

For the \ttH analysis case, three versions of the \ttH analysis were developed, with different computational characteristics that allowed to provide insight on how compute- and I/O-bound applications can be improved when using the framework. The three variations used of the \ttH analysis, \texttt{ttH\_as} (I/O-bound), \texttt{ttH\_sci} and \texttt{ttH\_scinp} (both compute-bound) showed performance improvements up to 6x, 15x, and 17x, respectively, over a conventional multiprocess approach for the same number of processes/threads.

HEP-Frame outperforms conventional strategies used for parallelization while removing the need of monitoring correct execution of hundreds of processes, which is often a time consuming task.
The improvements on performance, were consistent on 24-core Ivy Bridge and 32-core Broadwell servers. The performance of the Ivy Bridge server is similar to the Skylake when using a GPU device to automatically offload pseudo-random number generation for the \texttt{ttH\_sci} and \texttt{ttH\_scinp} applications.
HEP-Frame provided an overall speedup of 30x, 89x, 74x for the Ivy Bridge server with a Kepler GPU, and 31x, 258x, and 185x on the Intel Knights Landing manycore server, for \texttt{ttH\_as} (I/O-bound), \texttt{ttH\_sci}, and \texttt{ttH\_scinp}, respectively, over their original sequential implementations.

For the double and single top quark production at the HL-LHC, the analysis code was extended to several different final state topologies i.e., semileptonic and dileptonic $t\bar t$ production, as well as $t$-channel and $Wt$-channel associated production in the semileptonic channel, alone. Full kinematic reconstruction of events was performed by minimising a $\chi^2$ distribution, which allowed to classify events according to the number of top quarks and $W$-bosons, in the final state.

It was our intention to keep the physics case as simple as possible to justify the use of this tool. Recently, we have applied HEP-Frame in a global fit of $t\overline{t}$ and single top quark production considering all the contributions from the Standard Model processes, statistical uncertainties, and systematic uncertainties as well. In addition, the tool is able to preform event reconstruction without major difficulties. This global fit is intended to be preformed within the Standard Model Effective Field Theory (SMEFT). As expected, no difficulties nor major adaptations were necessary for a full analysis including statistical and systematic uncertainties.

\section*{Data Availability Statement}

Data sharing not applicable to this article as no datasets were generated or analysed during the current study.


\section{Appendix I}
\label{appendix_1}
This Appendix details an example of how to build a full analysis chain, based on the data file supplied with the package for download. The input file, corresponding to the generation of a $t\bar{t}H$ signal at the LHC, is provided after the DELPHES simulation of a typical LHC experiment, using the default ATLAS detector cards available. The input data file ($input\_event\_data.root$) is a ROOT file, with a $TTree*$ data structure named $Physics$, with several $Leaves$ that contain the events kinematic properties. Although any system could be used to execute the example, in what follows the CERN computing system was chosen, without any loss of generality. Upon login,
({\it ssh -l username lxplus.cern.ch}), the following global variables are defined (including a link to the BOOST library main directory),\\[1mm]
{\i
export dir   = main HEP-Frame source directory,\\
export data  = input data file directory,\\
export boost = Boost library main directory.
}\\[1mm]
\noindent
The $gcc$ compiler (in {\it \$GCC} directory) and ROOT versions (in {\it \$ROOT} directory) are set (if not done by default) with,\\[1mm]
\noindent {\it source \${GCC}/setup.sh}, \\
\noindent {\it source \${ROOT}/thisroot.sh}, \\
\noindent {\it export LD\_LIBRARY\_PATH=\${BOOST}/lib:\\
        \hspace*{4.4cm}\$LD\_LIBRARY\_PATH}\\[1mm]
\noindent As mentioned in Section~\ref{hep-frame}, the current version of the code was downloaded from,\\[1mm]
\textit{\url{https://bitbucket.org/ampereira/hep-frame/}},\\[1mm] unzipped (\textit{unzip download.zip}) and renamed to HepFrame
(\textit{mv ampereira-hep-frame HepFrame}).\\[-8mm]
\subsection{\bf Creating the Event Analysis}
\label{creating}
To create a new user analysis, the installation scripts (in the {\it scripts} directory) were run, using \\[1mm]
{\it cd HepFrame/scripts\\
./install.sh \$\{boost\}\\
./newAnalysis.sh AnalysisName \\ \hspace*{10mm}\$\{data\}/input\_event\_data.root Physics\\
cd ../Analysis/AnalysisName/src}\\[1mm]
where the following files can be found,\\[1mm]
\noindent {\it AnalysisName\_cfg.cxx\\
AnalysisName.cxx\\
AnalysisName\_EventBranches.h\\
AnalysisName\_Event.cxx\\
AnalysisName\_Event.h\\
AnalysisName.h\\
EventInterface.h}\\

\noindent 
Once available, the files can be edited and updated, according to user requirements.\\[-8mm]
\subsection{\bf Implementation and Running}
\label{running}
In order to implement the current analysis, few variables were included, as examples, in the variables list to be later saved for processing. These variables, $mTTH$ and $ptTTH$, correspond to the $t\bar{t}H$ mass and transverse momentum, respectively, were referred to be saved in {\it AnalysisName\_cfg.cxx},

{\small
\begin{verbatim}
#ifdef RecordVariables
mTTH
ptTTH
#endif
\end{verbatim}
}
\noindent and declared in {\it AnalysisName\_Event.h}
{\small
\begin{verbatim}
Float_t mTTH;
Float_t ptTTH;
\end{verbatim}
}
\noindent and initialised in {\it AnalysisName\_Event.cxx}
{\small
\begin{verbatim}
mTTH     =    0.0;
ptTTH    =    0.0;
\end{verbatim}}

\noindent To allow histograms to be saved into a ROOT file, useful after running, a new file (a TFile type) was declared in {\it AnalysisName.h} by inserting a new class variable and the histograms declaration, i.e.,

{\small
\begin{verbatim}
TFile *AnalysisName_Histos;
TH1D *hmTTH = new TH1D("hmTTH",  "ttH mass [GeV]",
                        100, 0., 300.);
TH1D *hptTTH = new TH1D("hptTTH", "ttH pt [GeV]",
                        100, 0., 300.);
\end{verbatim}}

\noindent Following all previous steps, the implementation of the event selection, analysis and histogram filling, was performed in {\it AnalysisName.cxx}, provided in this example. This is the final step before compiling and running.
The implemented code in the example resembles the following,\\[1mm]

{\small
\begin{verbatim}
void AnalysisName::finalize (void) {
    // Create a new file to store the  fChain TTree
    AnalysisName_Histos = new TFile("histos.root",
                                    "recreate");
    AnalysisName_Histos->cd();
    // output histograms              
    hmTTH->Write();
    hptTTH->Write();
    // close file
    AnalysisName_Histos->Close();
}

// Evaluate requirements for the specific cut
void cut_evaluation (unsigned this_event_counter) {
    // Put here your evaluations code
    TLorentzVector Top, Tbar, Higgs;
    // Initialization
    Top.SetPtEtaPhiM (PtTopQ, EtaTopQ,PhiTopQ, mTopQ );
    Tbar.SetPtEtaPhiM(PtTbarQ,EtaTbarQ,PhiTbarQ, mTbarQ);
    Top.SetPtEtaPhiM (PtHiggs,EtaHiggs,PhiHiggs, mHiggs);
    // ttH system 
    TLorentzVector ttH = Top + Tbar + Higgs;
    mTTH  = ttH.M();
    ptTTH = ttH.Pt();
}
// Sample cut, always define a cut like this
bool cut (unsigned this_event_counter) {
    // A cut should return true or false
    if (PtTopQ > 25.0 && abs(EtaTopQ) < 2.5 &&
        PtTbarQ > 25.0 && abs(EtaTbarQ) < 2.5 &&
        PtHiggs > 25.0 && abs(EtaHiggs) < 2.5) {
        cut_evaluation(this_event_counter);
        return true; 
    } else { return false;}
}
\end{verbatim}}

\noindent To compile the above code the user just needs to\\[-1mm]

\noindent{\it cd HepFrame/Analysis/AnalysisName\\
rm ../../lib/lib/libHEPFrame.a\\
make\\[-1mm]
}

\bibliographystyle{spphys}       
\bibliography{refs}

\end{document}